\documentclass[aps,epsf,eqsecnum,feynmp,graphics,amsfonts, amssymb, latexsym, amsmath,twocolumn]{revtex4}
\usepackage[latin1]{inputenc}
\usepackage{graphics}
\usepackage{amsfonts, amssymb, latexsym, amsmath}
\def\aut#1{#1}
\def\comment#1{}
\begin{document}
\title{Tunnelling Amplitudes
from Perturbation Expansions}
\author{B.~Hamprecht and H.~Kleinert}
\address{Institut für Theoretische Physik, Freie Universität Berlin,\\
Arnimallee 14, D-14195 Berlin, Germany\\
{\scriptsize e-mails: bodo.hamprecht@physik.fu-berlin.de}
{\scriptsize e-mails: hagen.kleinert@physik.fu-berlin.de}}
\begin{abstract}
We present a method for
extracting tunnelling amplitudes
from
perturbation expansions
which are  always
divergent
and not Borel-summable.
We show that
they can be  evaluated
by
 an analytic continuation
 of variational perturbation
 theory. The power of the method
is illustrated
by calculating the
imaginary parts
of the partition function
of the anharmonic oscillator
in zero spacetime dimensions
and of
the
 ground state energy of
 the anharmonic oscillator for all negative
 values of
 the coupling constant $g$
and show that they are in excellent agreement with the exactly
known
values.
As a
highlight
 of the theory
we recover
from the divergent
perturbation expansion of the tunnelling amplitude
the action of the instanton and
 the effects of higher  loop
fluctuations
around it.
\end{abstract}
\maketitle
\section{Introduction}
\noindent{\bf 1.}
Tunnelling processes
govern
many
important physical phenomena.
Their theoretical description
requires the
calculation of
the contribution of
critical bubbles
to the partition function,
including
their
fluctuation entropy \cite{Langer}.
The latter can be found at
most to
one-loop order.
Higher loop effects
are
prohibitively
complicated \cite{Hagen}.
It would be of great advantage
tunnelling amplitudes
 could be derived
from  ordinary
perturbation expansions
around the free theory
since these can be performed
to many loops
\cite{Verena}.
The difficulty
arising in such a program
is that tunnelling amplitudes are described by the analytic continuation
of divergent Borel-summable
power series expansions
in the coupling constant $g$
to negative $g$ where they become
non-Borel-summable.
None of
 the currently
known resummation schemes \cite{Verena,Resumm}
is able to deal with such expansions.
Some time ago it was suggested that
a resummation is possible
by variational perturbation theory \cite{Karr}.
However, the imaginary parts calculated there
gave accurate
imaginary parts  only in the {\em sliding regime\/}
of larger negative $g$
and
did not
invade into the
proper tunnelling regime of
small $g$ dominated by critical bubbles.
A separate variational treatment
of an instanton calculation
was set up to cover this region \cite{Tunn}.

In this note we show that
 the
non-Borel-summable
series  can be evaluated with any desired precision
 by an appropriate
continuation of  variational perturbation
 theory \cite{Hagen,Verena} in
 to
 the complex $g$-plane.

Variational perturbation
 theory
has a long history
\cite{refs,finiteg,KleinertJanke,Guida}. It is based on the introduction of a
dummy variational
 parameter $ \Omega $
on which
the full perturbation expansion does not depend,
while the truncated expansion
does. An optimal $ \Omega $ is then selected
by the principle of minimal sensitivity \cite{Stevenson},
 requiring the
quantity of interest to be stationary as function of the
variational parameter. The optimal
$ \Omega $
 is usually
taken from a zero of
 the derivative with
respect to $ \Omega $.
If the first derivative has no zero,
a zero of the second derivative is chosen.
For Borel-summable series, these zeros are always real, in contrast
to statements
in
the literature
 \cite{Neveu,RULES,Braaten,Ramos}
advocating
the use of complex
zeros. Complex zeros, however, produce
in general  wrong results
for Borel-summable series, as was recently demonstrated
in Ref.~\cite{AntiBraaten}.

The purpose of this note is to show that there
does exist
 a wide range of applications
of complex zeros in the
previously untreatable
field of non-Borel-summable series.
These arise typically
in tunnelling problems, and we
shall
see
 that variational perturbation
theory provides us with an
efficient method for evaluating
these series and rendering
their real and imaginary parts
with any desirable accuracy
if only enough perturbation coefficients are available.
The choice of the
 complex zeros
is dictated by the
requirement
to achieve at
each order
the least oscillating
imaginary part when approaching
 the tip of
 the cut. We shall call this selection rule
the {\em principle of minimal
sensitivity and oscillations\/}.

\noindent{\bf 2.}
 For an introduction
to the method
consider the exactly known partition function
of an  anharmonic oscillator
in zero spacetime dimensions,
 which is a simple integral
representation of a modified  Bessel function $K_ \nu (z)$:
\begin{eqnarray}
\label{FOKKER}
Z(g) &=& \frac{1}{\sqrt{2\pi}}
\int_{-\infty}^\infty dx\,\exp{(-x^2/2-g~x^4/4)}~~~~~~~~
 \nonumber \\
&=&{e^{1/8g}}{(4\pi g)^{-1/2}}~{K_{1/4}(1/8g)}~,~~~~~~~~~
\end{eqnarray}
For small $g<0$,
 the function $Z(g)$ and
its inverse $D(g)\equiv Z^{-1}(g)$ have a divergent non-Borel-summable
 power
series:
\begin{equation}
D(g)=  \sum_{l=0}^\infty
 a_lg^l,
\label{@}\end{equation}
In the
strong-coupling regime, there exists a convergent expansion
\begin{align}
\label{FP-STRONG}
D(g) = g^ \alpha \, \sum_{l=0}^\infty\; b_l\;g^{-\omega l},
\end{align}
with $ \alpha =1/4$ and
$\omega=1/2$. In the context of critical phenomena,
the exponent
$ \omega $ coincides with the the  Wegner exponent \cite{Wegner}
of approach to scaling \cite{strong}.
The $L$th variational approximation
depending on the variational parameter $ \Omega $
is given by the truncated series
\cite{Hagen,Verena}
\begin{align}
\label{FP-VAR}
D_{\rm var}^{(L)}(g,\Omega) = \Omega^{p} ~\sum_{j=0}^L \left(\frac{g}{\Omega^q}\right)^j \epsilon_j(\sigma)\,,
\end{align}
where $q=2/\omega=4$, $p= \alpha q=1$.
Introducing the
parameter
$\sigma=\Omega^{q-2}(\Omega^2-1)/g$, the re-expansion coefficients are
\begin{align}
\label{FP-EPS}
\epsilon_j(\sigma) = \sum_{l=0}^j a_l \binom{(p-lq)/2}{j-l} (-\sigma)^{j-l}\,.
\end{align}
 Following
 the principle of minimal sensitivity,
we have to find  the zeros of
 the derivative of $\partial _ \Omega D_{\rm var}^{(L)}(g,\Omega)$,
which happen to coincide with the
zeros of a function of the variable $ \sigma $
only, to be denoted by $\zeta ^{(L)}(\sigma)$:
\begin{align}
\label{FP-DERIV}
 \zeta ^{(L)}(\sigma)\! = \!\sum_{l=0}^L a_l\,
 \binom{(p\!-\!lq)/2}{L\!-\!l}
(p\!-\!lq\!+\!2l\!-\!2L)
 (-\sigma)^{L-l}\,.
\end{align}
For a proof of this
remarkable property
see
 \cite{JKsig},
and
 the textbook  \cite{Hagen}
(p. 291).

\begin{figure}[htp!]
\begin{center}
\setlength{\unitlength}{1cm}
\begin{picture}(18.5,12)
\put(0.5,6.4){\scalebox{.8}[.8]{\includegraphics*{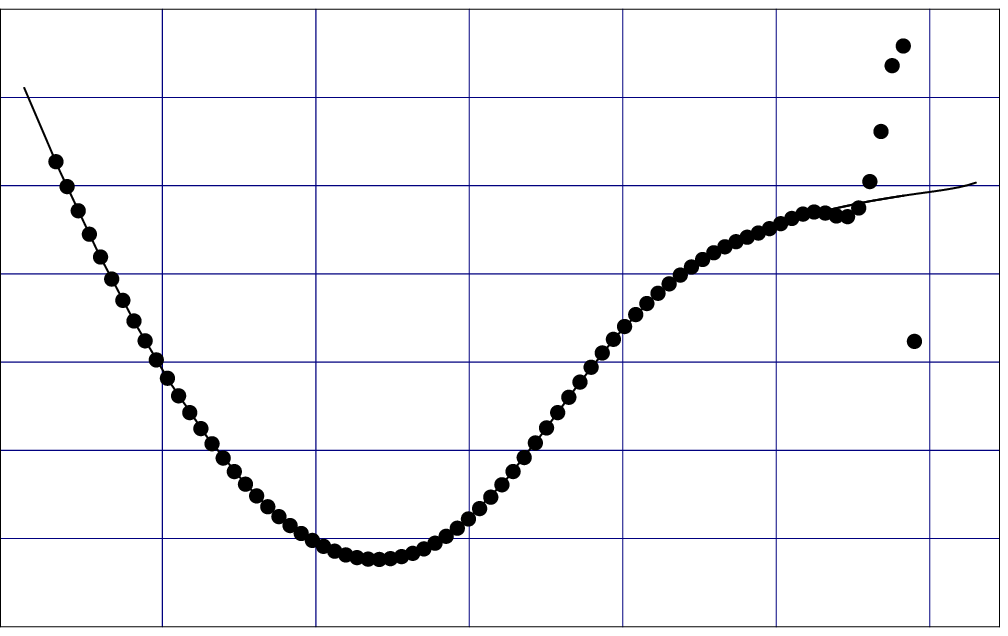}}}
\put(0.5,.4){\scalebox{.8}[.8]{\includegraphics*{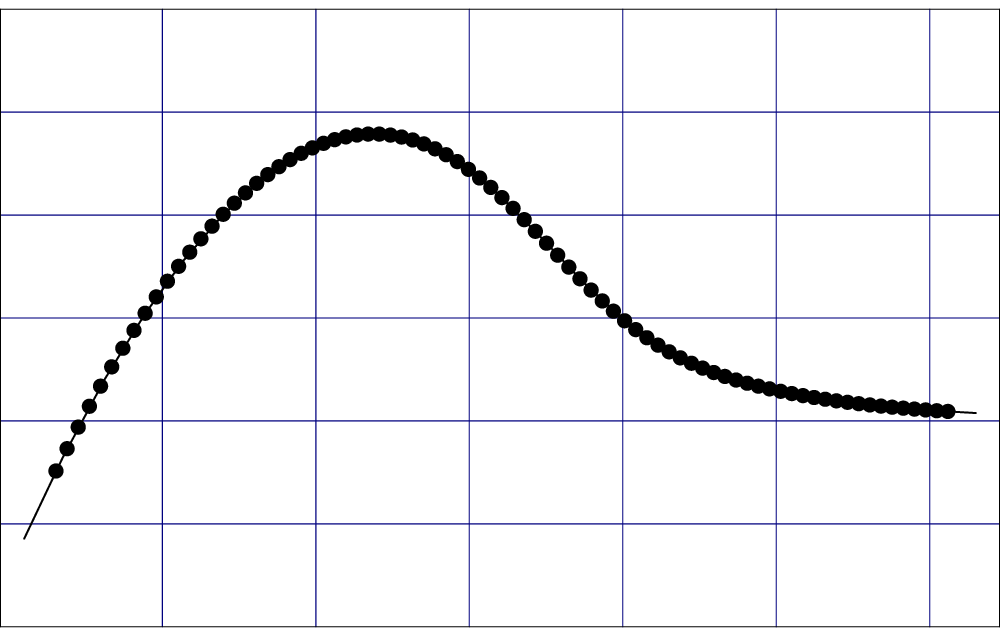}}}
\put(3.1,10.3){\small{Im $D(g)\exp{(-1/4g)}$}}
\put(3.1,1.6){\small{Re $D(g)$}}
\put(.2,9.9){$\small{.7}$}
\put(.2,8.5){$\small{.6}$}
\put(.2,7.1){$\small{.5}$}
\put(-.05,3.7){$\small{-.9}$}
\put(-.05,2.){$\small{-1.}$}
\put(1.7,6.1){$\small{0}$}
\put(4.,6.1){$\small{-2}$}
\put(6.55,6.1){$\small{-4}$}
\put(1.7,0.1){$\small{0}$}
\put(4.,0.1){$\small{-2}$}
\put(6.55,0.1){$\small{-4}$}
\put(8,6.1){$\log(- g)$}
\put(8,.1){$\log(-g)$}
\end{picture}
\caption[NB-IV]{Imaginary and real parts
from variational perturbation theory
of 16th order $D_{\rm var}^{(16)}(g)$ as a function of $\log{(-g)}$
(dots)
in comparison with exact curves (curves).
In the imaginary part
we have removed
the leading
essential singularity
by dividing out a  factor  $\exp{(1/4g)}$
to see the amazing accuracy
with which this singularity is approximated. For very small $g$ the
onset of oscillations in the imaginary part can be seen
which moves
towards the origin for increasing order
$L$ of the variational approximation.
 }
\label{NB-IV}
\end{center}
\end{figure}

As an example,
we take
 the weak-coupling expansion
to $L=16$th order and
calculate real and
imaginary parts for
 the non-Borel region $-2<g<-0.008$
selecting the zero
of $ \zeta ^{(16)}(\sigma)$
according to the principle
of minimal sensitivity and
oscillations.
The result is
 shown in
 Fig.~\ref{NB-IV}.
In order to point out
how well
the variational result
approximates  the
essential singularity
of the imaginary part
 $ \propto
\exp(-1/4g)$ for small $g<0$,
 we have removed this factor.
The agreement is excellent down to very small
$-g$.

\begin{figure}[htp!]
\begin{center}
\setlength{\unitlength}{1cm}
\begin{picture}(15,8)
\put(.5,.4){\scalebox{.8}[.8]{\includegraphics*{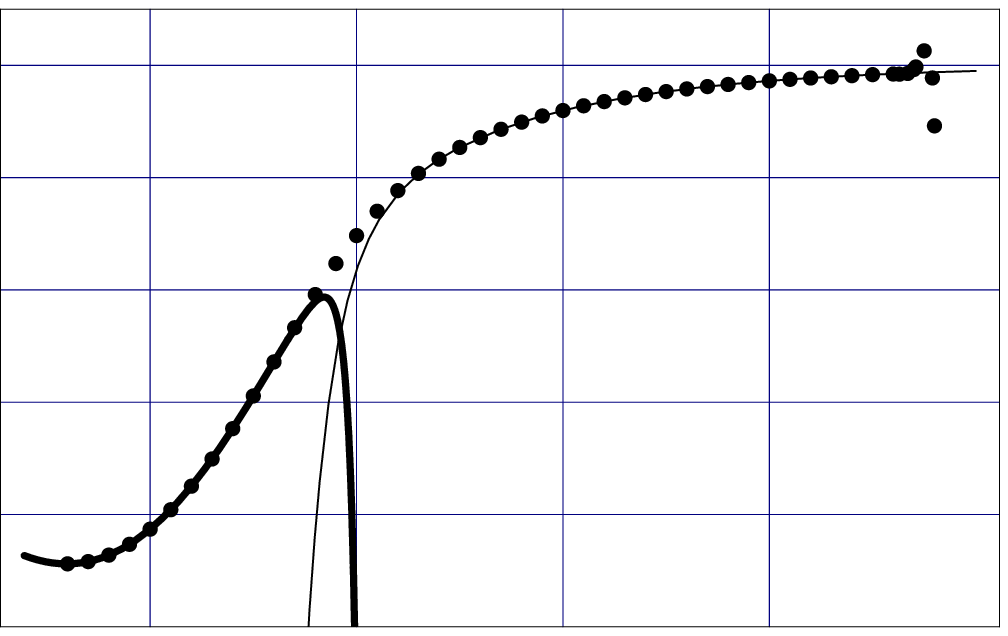}}}
\put(-.1,1.25){$-.8$}
\put(-.1,3.1){$-.4$}
\put(5.6,3.5){$l(g)$}
\put(.15,4.9){$0$}
\put(1.43,.1){$-2$}
\put(3.1,.1){$-3$}
\put(4.8,.1){$-4$}
\put(6.45,.1){$-5$}
\put(3.65,-.25){$\log(-g)$}
\end{picture}
\caption[I]{Logarithm of
 the imaginary part of
 the ground state energy of
 the anharmonic oscillator with the essential singularity
factored out for better visualization,
$
l(g)=\log\left[ {\sqrt{-\pi g/2}~E_{0,\rm var}^{(64)}(g)}\right] -1/3g$,
plotted against
 small negative values of
 the coupling constant $-0.2<g<-.006$
where the series is
 Borel-nonsummable.
The thin curve represents
the divergent expansion around an instanton
of Ref.~\cite{ZINNJ}.
The fat curve is the
 $22$nd order approximation of the
 strong-coupling expansion,
 analytically continued
to negative $g$ in the sliding regime calculated
 in
Chapter 17 of the textbook \cite{Hagen}.
 }
\label{I}
\end{center}
\end{figure}

\noindent{\bf 3.}
 Let us now turn to the nontrivial problem
of summing the
 instanton region of
 the anharmonic oscillator for $g<0$.
The divergent weak-coupling perturbation expansion for
 the ground state energy of
 the anharmonic oscillator
 with
 the potential $V(x)=x^2/2+g\,x^4$ is, to order $L$;
\begin{align}
\label{WEAK}
E_{0,\rm weak}^{(L)}(g) = \sum_{l=0}^L\; a_l\;g^l\,,
\end{align}
where $a_l=(1/2,~3/4,~-21/8,~333/16,~-30885/128,~\dots)$.
The expansion is obviously
not Borel-summable
 for $g<0$, but will now be evaluated
with our new technique,
proceeding in the
 same way as
for the above test function $D(g)$
via
Eqs.~(\ref{FP-VAR}) through (\ref{FP-DERIV}).
The known strong-coupling growth parameters
are
 $ \alpha =1/3$ and
$\omega=2/3$, so that
$p=1$ and
 $q=3$
 in Eq.~(\ref{FP-DERIV}) which will guarantee
 the correct scaling properties for $g \to \infty$.
To order $L=64$ we obtain
from
 the optimal  zero
of $ \zeta ^{(64)}( \sigma )$
 the logarithm of
 the scaled imaginary part
\begin{align}
\label{IM64}
l(g):=
\log\left[ {\sqrt{-\pi g/2}~E_{0,\rm var}^{(64)}(g)}\right] -1/3g\,,
\end{align}
 shown
 in Fig.~\ref{I} for $-0.2<g<-0.006$. The point $g=-0.006$ is
 the closest approach to
 the singularity at $g=0$ for $L=64$ before
 the onset of oscillations.

Let us compare
 our curve with
the expansion
derived
 from
 instanton calculations \cite{ZINNJ}:
\begin{align}
\label{ZJ}
f(g)&=b_1 g-b_2 g^2+b_3 g^3-b_4 g^4+\dots ~,
\end{align}
with
coefficients
$b_1 =3.95833,~b_2=19.344,~b_3=174.21,~b_4=2177$.
This
expansion is divergent and
non-Borel-summable for $g<0$.
Remarkably, we are able to extract this expansion
from our data points.
Since our result is given by a convergent
expansion, the fitting procedure will
 depend somewhat on
 the
 interval chosen over which we fit
our curve by a power series.
A compromise between a sufficiently long
 interval and
the runaway of
 the divergent
 instanton expansion is obtained for
a lower limit
 $g>-.0229\pm .0003$
and an
 upper limit $g=-0.006$. Fitting a polynomial to
 the data, we extract
 the following
 first three coefficients:
\begin{eqnarray}
\label{INST}\!\!\!\!
b_1 \!=\!3.9586\pm .0003,~b_2\!=\!19.4\pm .12,~b_3\!=\!135\pm 18.
\label{@agree}\end{eqnarray}

\noindent{\bf 4.}
The agreement
of our curves
 with the exact results
 in
Figs.~\ref{NB-IV} and
of our
 expansion coefficients
in (\ref{@agree}) with the exact ones in (\ref{ZJ})
demonstrates
 that
our method is capable of probing
 deeply into
 the
 instanton region of the coupling constant.

Let us end by remarking that
another procedure
of summing non-Borel series can be deduced
from
a development
in the first of Refs.~\cite{strong}
(see also
Chapter 20
of the textbook \cite{Verena})
One may
 derive
 a strong-coupling
expansion of the type
(\ref{FP-STRONG}) from the divergent weak-coupling expansion,
which can then
be
continued
analytically
 to negative $g$ by a simple rotation of
the power
$g^{- \omega l}$ to
$e^{-i\pi  \omega l}(-g)^{- \omega l}$.
This method is  applicable
only
in the sliding regime.
In Fig.~\ref{I},
we have plotted the resulting curve
 to order
$L=9$.
The present
work fills the gap
between the sliding regime
and the tunnelling regime
by extending variational perturbation theory
to all $g$ arbitrarily close to zero, without the need
 for a separate treatment of the tunnelling regime.

There exists, of course,
  a wealth of
 possible applications  of this
simple theory, in particular to quantum field theory
where variational perturbation theory has so far yielded
the most accurate
critical exponents
from Borel-summable
series \cite{strong,Verena,Lipa}.

~\\Acknowledgment\\
\\[-.9em]
This work was partially supported by
ESF COSLAB Program and by the Deutsche Forschungsgemeinschaft
under Grant Kl-256.

\end{document}